\documentclass[doublecol]{epl2}
\usepackage{amssymb}
 
\title{Equivalence of the Klein-Gordon random field and the complex Klein-Gordon quantum field}
\shorttitle{Equivalence of Klein-Gordon random/quantum fields}
\author{P. Morgan}
\institute{Physics Department, Yale University, New Haven, CT 06520, USA.}

\abstract{
The difference between a Klein-Gordon random field and the complex Klein-Gordon quantum field
is characterized, explicitly comparing the roles played by negative frequency modes
of test functions in creation and annihilation operator presentations of the two theories.
The random field and the complex quantum field can both be constructed from the same
creation and annihilation operator algebra, making them equivalent in that sense.
}

\newcommand\Half{{\frac{1}{2}}}
\newcommand\SHalf{{\scriptstyle\frac{1}{2}}}
\newcommand{\HSPACE}[1]{{\hspace{#1}}}    

\newcommand{\AO}{{\mathring{a}}}
\newcommand{\AP}{{a}}
\newcommand{\A}{{\mathbf{a}}}
\newcommand{\B}{{\mathbf{b}}}
\newcommand{\AZ}{{\mathsf{a}}}

\begin{document}
\maketitle

\section{Introduction}
The ``Klein-Gordon random field'' is taken here to be a commutative quantum field for which
$[\hat\phi(x),\hat\phi(y)]=0$ for all $x$ and $y$, satisfying microcausality trivially.
The random field structure is of interest partly because it admits a Lie field deformation that
preserves commutativity of the random field~\cite{MorganJMP}, whereas there is a no-go theorem
proving that Lie field deformations of Wightman fields that preserve nontrivial microcausality
are not possible~\cite{Baumann}.
{
A state over a random field can be presented in more directly probabilistic ways, but  it is advantageous to give an
algebraic and Hilbert space presentation of both when the aim is
to show how closely a random field model may parallel a quantum field model.
A discussion of the mathematics of random fields in the quantum field context may be found in~\cite{Gudder}.
A selection of approaches that are more-or-less in terms of random fields is listed in~\cite{Khrennikov}, to
which may be added~\cite{Elze,Wetterich,Helland}.
An argument that Bell inequalities are generally not satisfied by random fields may be found
in~\cite{MorganJPA}.}

A relatively abstract comparison of the Klein-Gordon random field with the complex Klein-Gordon quantum field
shows that both fields involve negative-frequency modes of test functions.
Negative-frequency modes have generally been understood as positive-frequency antimatter
modes~{\cite{BH,Wallace,Saunders}},
however we here engage with the algebraic structure in a way that clarifies the parallel with random fields.
{
The distinction between positive- and negative-frequency modes is somewhat problematic for quantum fields because
it is not well-defined in curved space-times and for accelerating observers~\cite{BH}, whereas we will see that
there is no need for a distinction between positive- and negative-frequency modes in the more natural mathematics
of random fields.}

Part of the motivation for this Letter is that we might, in time, use Lie random fields to construct models
for \emph{experiments}, following the principles of Bell's polemic~\cite{Bell}, in contrast to accepting the
focus of quantum theory on constructing models for measurement and preparation apparatuses that in principle
are \emph{not perfectly} separable in the context of a given experiment (see also~\cite{MorganJPA,MorganJMP}).
{
The idea that measurement apparatuses should be modeled explicitly as part of models of experiments is also
expressed by Feynman \& Hibbs, ``The usual separation of observer and observed which is now needed in
analyzing measurements in quantum mechanics should not really be necessary, or at least should
be even more thoroughly analyzed. What seems to be needed is the statistical mechanics of amplifying
apparatus.''~\cite[pp22-23]{FH}; such a model constructed in a quantum mechanical formalism may be found,
for example, in ~\cite{ABN}.
A detailed thermodynamics of measurement apparatuses is also required if we take seriously the insistence
of the Copenhagen interpretation, which has recently been given fresh life by~\cite{Landsman}, that we should
give a classical description of an experimental apparatus that is sufficient for us to reproduce experimental
results --- in this context, a thermodynamic or kinetic theory model of the preparation and measurement apparatus
and the raw measurement results is required to be classical.
As we introduce classical models of increasing detail for an experiment, we effectively move the Heisenberg
cut to smaller scales, in contrast to the more common approach that moves the Heisenberg cut to larger
scales to include more of an experimental apparatus in the quantum model.}

{
When an experimental apparatus is considered as a whole --- instead of making an ad hoc separation into
preparation apparatus and measurement apparatus or, less operationally, into measured systems and measurement
apparatus --- the world-tubes of all parts of the experimental apparatus are time-like separated from the past and
future of the world-tubes of all other parts.
If we associate a measurement operator with a measurement apparatus, which is a classical object that we can
control, instead of with an individual measurement event, which is a thermodynamic transition the timing of which
we cannot control, then the idea of microcausality as a necessary guiding feature of quantum field theory is worrying,
because the world-tubes of measurement apparatuses that are part of a given experiment are all time-like separated
from one another.
It is problematic, and dependent on what interpretation we prefer, to associate a measurement operator with a single
event instead of with a measurement apparatus because empirical verification is by comparison of statistics of
experimentally observed properties of measurement events with probability densities generated using a state and an
operator. There is no direct parallel of individual events in the mathematics of quantum theory.}

{
A common extreme considers the whole universe, not just a whole experiment, to be modeled by a quantum state, with
measurement having a metaphysical status not associated with real measurement apparatuses.
A measurement apparatus is modeled by parts of the state instead of by a measurement operator.
For a quantum state that models the whole universe, idealized measurements are not associated with measurement apparatuses,
so they are not subject to empirical verification, so we may assign whatever commutation relations are convenient.
Quantum theory in this extreme ceases to be a model of measurement, contrary to the original positivist principles.}

{
A non-commuting algebra of observables is a very effective mathematical model for the first-order effects of quantum
fluctuations on measurement, but interactions more generally are not modeled in a consistent way in conventional
quantum field theory.
Instead, algebraic structure is used to model the effects of quantum fluctuations, while other effects are modeled
by terms in a Lagrangian or in a Hamiltonian.
The dynamical and thermodynamic relationships between parts of an experimental apparatus, which depend on the precise
structure and properties of the whole apparatus, should be modeled consistently, as far as possible.
The suggestion of~\cite{MorganJMP} is to model all interactions using algebraic structure, which closely parallels
algebraic quantum field theory and deliberately leaves questions of dynamics unaddressed, but of course we might
prefer to model all interactions using a random field dynamics.}

{
A complete model of an experimental apparatus is relatively intractable, so we consider now the way in which
experiments are modeled in practice.
}
We generally require quantum fields to be complex linear maps from a space of complex-valued test
functions into a space of operators that satisfy microcausality.
We are also accustomed, however, to using creation and annihilation operators and the vacuum projection
operator $\left|0\right>\left<0\right|$ quite freely to model measurements, particularly in particle physics
and in quantum optics, even though the larger algebra does not satisfy microcausality{, because
projection operators allow us to construct simple models for yes/no and other discrete-valued experimental data.}
Much of the empirical success of quantum field theory is in terms of the larger algebra.
If the algebra of creation and annihilation operators is taken to be empirically supported, then this equally
supports understanding the mathematics in terms of a Klein-Gordon random field or in terms of a complex
Klein-Gordon quantum field, given that, as we will show below, either can be constructed from the
creation and annihilation operator algebra of the other.

\section{The Klein-Gordon random field}\label{KGrf}
The Klein-Gordon random field can be presented relatively abstractly as a complex-linear map from a Schwartz
space of complex-valued test functions $\mathcal{S}$ into an abstract $\star$-algebra that is freely generated
by unbounded creation and annihilation operators,
\begin{equation}
  \hat\phi:\mathcal{S}\rightarrow\mathcal{A};\HSPACE{0.5em}\hat\phi_f=\AO_{f^*}+\AO^\dagger_f, \qquad [\AO_f,\AO_g]=0,
\end{equation}
\begin{eqnarray}\label{GoodA}
  [\AO_f,\AO_g^\dagger]&=&(f,g)\cr
    &=&\hbar\int\tilde f^*(k)\tilde g(k)2\pi\delta(k^\mu k_\mu-m^2)\frac{\upd^4k}{(2\pi)^4}.
\end{eqnarray}
The test function conjugation $f\mapsto f^*$ is a local operation in real space, $f(x)\mapsto f^*(x)$, so that
$\widetilde{f^*}(k)=\tilde f^*(-k)$ in Fourier space; $(f,g)$ is a Poincar\'e invariant inner product.
Note that we adopt here an opposite convention from that of~\cite{MorganJMP}, in that $\AO_f^\dagger$ is taken
to be complex linear, so that $\AO_f$ is complex antilinear.
It is paramount that the definition of the algebra includes positive and negative frequencies equally,
so that $[\hat\phi_f,\hat\phi_g]=(f^*,g)-(g^*,f)=0$.

In a Fourier space presentation, giving up the useful isolation of space-time structure in the inner
product $(f,g)$ on the test function space, we can write
\begin{eqnarray}
  &&\hat\phi(x)=\int\left[\AO(k)e^{-ik_\mu x^\mu}+\AO^\dagger(k)e^{ik_\mu x^\mu}\right]\frac{\upd^4k}{(2\pi)^4},\\
  &&\HSPACE{-2em}
    \left[\AO(k),\AO^\dagger(k')\right]=\hbar(2\pi)^4\delta^4(k-k')2\pi\delta(k^\mu k_\mu-m^2).\label{NaturalUnsignedkk}
\end{eqnarray}
It is common to see such constructions with different, frequency-dependent normalizations of the creation and
annihilation operators, however the above choice is a natural manifestly Lorentz invariant normalization.

The operator $\hat\phi_f$ is a self-adjoint observable $\hat\phi_f^\dagger=\hat\phi_{f^*}=\hat\phi_f$ only if
$f=f^*$ is real-valued; we can trivially construct a self-adjoint observable for any complex test function $f$,
\begin{equation}
  \hat R_f=\SHalf(\hat\phi_f+(\hat\phi_f)^\dagger)
          =\SHalf(\AO_f+\AO_{f^*}+\AO_f^\dagger+\AO_{f^*}^\dagger),
\end{equation}
from which we can recover $\hat\phi_f=\hat R_f-i\hat R_{if}$.
The observables $\hat R_f$ satisfy the trivial commutation relation $[\hat R_f,\hat R_g]=0$ for all test
functions.

The vacuum vector $\left|0\right>$ is defined by the operation $\AO_f\left|0\right>=0$ of annihilation operators
acting on the vacuum vector and the normalization $\left<0\right|\left.\!0\right>=1$, which allows us to use the
Gelfand-Naimark-Segal (GNS) construction~(see, for example,~\cite[\S III.2.2]{Haag}) of a Hilbert space because
$A\mapsto\left<0\right|A\left|0\right>$ is a state over the $\star$-algebra of creation and annihilation operators.

The Klein-Gordon random field algebra can also be presented using independent annihilation operators
$\AP_f$ and $b_f$,
\begin{equation}
 \hat\phi:\mathcal{S}\rightarrow\mathcal{A};\HSPACE{0.5em}\hat\phi_f=\AP_{f^*}+\AP^\dagger_f+b_{f^*}+b^\dagger_f,
\end{equation}
$[\AP_f,\AP_g]=[\AP_f,b_g]=[\AP_f,b_g^\dagger]=[b_f,b_g]=0$,
\begin{eqnarray}\label{GoodAB}
  [\AP_f,\AP_g^\dagger]&=&(f,g)_+\cr
  [b_f,b_g^\dagger]&=&(f,g)_-=(g^*,f^*)_+\cr
    (f,g)_+&=&\HSPACE{-0.75em}\hbar\HSPACE{-0.3em}\int\HSPACE{-0.3em}
       \tilde f^*(k)\tilde g(k)2\pi\delta(k^\mu k_\mu-m^2)\theta(k_0)\frac{\upd^4k}{(2\pi)^4},\cr
    (f,g)_-&=&\HSPACE{-0.75em}\hbar\HSPACE{-0.3em}\int\HSPACE{-0.3em}
       \tilde f^*(k)\tilde g(k)2\pi\delta(k^\mu k_\mu-m^2)\theta(-k_0)\frac{\upd^4k}{(2\pi)^4}.
\end{eqnarray}
We can use $\AO_f=\AP_f+b_f$ whenever it is convenient to do so.
Note that the Klein-Gordon random field does not require a separation into positive- and negative-frequency
modes, but we can introduce the distinction if we wish.

\section{The complex Klein-Gordon quantum field}\label{cKGqf}
We can present the complex Klein-Gordon quantum field in similarly abstract terms, as
\begin{equation}
  \hat\psi_f=\A_{f^*}+\B_f^\dagger,
\end{equation}
$[\A_f,\A_g]=[\A_f,\B_g]=[\A_f,\B_g^\dagger]=[\B_f,\B_g]=0$,
\begin{eqnarray}\label{GoodQAB}
  [\A_f,\A_g^\dagger]&=&(f,g)_+,\cr
  [\B_f,\B_g^\dagger]&=&(f,g)_+,
\end{eqnarray}
so that $[\hat\psi_f,\hat\psi_g]=0$, $[\hat\psi_f,(\hat\psi_g)^\dagger]=(f^*,g^*)_+ -(g,f)_+$.
$\hat\psi_f$ is never a self-adjoint observable, so we construct the self-adjoint observables
\begin{equation}
  \hat O_f=\hat\psi_f+(\hat\psi_f)^\dagger=\A_{f^*}+\A_{f^*}^\dagger+\B_f+\B_f^\dagger,
\end{equation}
in parallel with the construction of $\hat R_f$, from which we can recover
$\hat\psi_f=(\hat O_f-i\hat O_{if})/2$, and for which we have the commutation relations
\begin{eqnarray}
  \left[\hat O_f,\hat O_g\right]
      &=&\HSPACE{-.85em}
         \bigl((f,g)_+ -(g,f)_+\bigr) - \bigl((g^*,f^*)_+ -(f^*,g^*)_+\bigr)\cr
      &=&\HSPACE{-.85em}
         \bigl((f,g)_+ -(g,f)_+\bigr) - \bigl((f,g)_- -(g,f)_-\bigr)\cr
      &=&\HSPACE{-.85em}
         \bigl((f,g)_+ +(g,f)_-\bigr) - \bigl((g,f)_+ +(f,g)_-\bigr).
\end{eqnarray}
We observe that there are equivalent but oppositely signed contributions to the commutation relations for
positive- and negative-frequency modes of the test functions $f$ and $g$.
Microcausality is satisfied, $[\hat O_f,\hat O_g]=0$, whenever the supports of $f$ and $g$ are space-like
separated, because in that case $(f,g)_+=(g^*,f^*)_+=(f,g)_-$.

We can construct $\hat O_f$ as a sum of creation and annihilation operators,
$\hat O_f=\AZ_f+\AZ_f^\dagger$, where $\AZ_f=\A_{f^*}+\B_f$ satisfies the commutation relation
\begin{eqnarray}\label{AlternateKGqf}
  [\AZ_f,\AZ_g^\dagger]&=&(f,g)_+ +(f^*,g^*)_+=(f,g)_+ + (g,f)_-\cr
    &=&\HSPACE{-0.75em}\hbar\HSPACE{-0.3em}\int\HSPACE{-0.3em}
       \left[\tilde f^*(k)\tilde g(k)+\tilde f(-k)\tilde g^*(-k)\right]\cr
    &&\HSPACE{2.25em}\times\HSPACE{0.5em}2\pi\delta(k^\mu k_\mu-m^2)\theta(k_0)\frac{\upd^4k}{(2\pi)^4}\cr
    &=&\HSPACE{-0.75em}\hbar\HSPACE{-0.3em}\int\HSPACE{-0.3em}
       \widetilde{f^\bullet}^*(k)\widetilde{g^\bullet}(k)
                 2\pi\delta(k^\mu k_\mu-m^2)\frac{\upd^4k}{(2\pi)^4}\cr
    &=&(f^\bullet,g^\bullet),
\end{eqnarray}
where
\begin{equation}
  \widetilde{f^\bullet}(k)=\theta(k_0)\tilde f(k) + \theta(-k_0)\tilde f^*(k)
\end{equation}
applies complex conjugation only to negative frequency components.
Such a construction violates the spirit of a somewhat implicit axiom that quantum fields must be complex
linear functionals of test functions, but is equivalent to the conventional construction.
{
The nontrivial complex structure required to construct the complex Klein-Gordon quantum field is especially
noted in \cite{BH}.
}

\section{Discussion}
It is immediate from the above that the algebras of creation and annihilation operators are isomorphic,
$\AZ_f\mapsto\AO_{f^\bullet}$, $\AO_f\mapsto\AZ_{f^\bullet}$, so we can construct either the Klein-Gordon
random field or the complex Klein-Gordon quantum field, both of which satisfy microcausality, if we are
given either algebra of creation and annihilation operators.
Hence, from the operators $\hat R_f=\Half(\AZ_{(f+f^*)^\bullet}+\AZ_{(f+f^*)^\bullet}^\dagger)$ and
$\hat O_f=\AO_{f^\bullet}+\AO_{f^\bullet}^\dagger$, we can reconstruct the Klein-Gordon random field
and the complex Klein-Gordon quantum field {in each algebra}.
We note that the operation $f\mapsto f^\bullet$ can be understood to be local in this context.

{
The isomorphism $\AZ_f\mapsto\AO_{f^\bullet}$ is a strange admixture of unitary and anti-unitary equivalence,
however it is enough to allow us to construct a Klein-Gordon random field model that has the same phenomenology
as any given complex Klein-Gordon quantum field model.
Insofar as experiments are modeled by projection operators constructed using the vacuum projector and creation
and annihilation operators, we can construct random and quantum field models that predict identical experimental
results, using the slightly different test functions appropriate to the different models.
$\AZ_f$ and $\AO_f$ are empirically equally capable if we consider that the test functions to be used when
constructing an empirically adequate model for a given measurement or preparation apparatus are determined by
experimental data; test functions are not given \textit{a priori}.}
It is of course the case that the Klein-Gordon random field for positive-frequency modes is identical to the
complex Klein-Gordon quantum field for the same positive-frequency modes, it is only for negative frequencies
that the random and complex quantum fields are distinct.

{
A difference arises, however, if we consider what experimental results we would obtain if we were to
measure either $\hat\phi_f$ or $\hat\psi_f$, especially because it is always possible to make joint measurements
of many random field observables, whereas joint measurements are not always possible for a quantum field.
The scale of the incompatibility between quantum field measurements is determined by the scale of Planck's
constant, which also determines the scale of quantum fluctuations~\cite{MorganPLA}, whereas there is no
incompatibility between random field measurements.
If we regard quantum field and random field observables both as models for ideal measurements, neither of which can
be implemented precisely, but in terms of which we can construct models for real experimental apparatuses, then there
is no necessity to make a choice of one or of the other.
We can use whatever operators we find useful from the whole algebra that can be constructed using creation and
annihilation operators and the vacuum projection operator to model experimental results.
Insofar as quantum fluctuations and the effects of quantum fluctuations on measurement are universal and absolutely
constant through all space-time (apparently very differently from thermal fluctuations), it is likely simpler to
consider the quantum field to be a fundamental ideal measurement, but we can still discuss what measurement results
we \emph{would} obtain \emph{if} we did have such a measurement apparatus even if we cannot find a measurement apparatus
that implements measurements of the random field observable $\hat\phi_f$.
}

The Klein-Gordon random field goes somewhat against the standard idea that only positive frequencies
are permitted, but there are several reasons why we should not insist on positive frequency, which in
quantum theory is generally taken to equate to positive energy.
First, we note that the complex Klein-Gordon quantum field has insinuated negative frequency modes
by a judicious use of complex conjugation, which undermines a restriction to positive frequency.
Second, ``stability'' is frequently said to be why energy must be bounded below, but, by analogy
with thermal states, the vacuum state can be \emph{thermodynamically} stable without being the lowest
energy state, by being the most Poincar\'e invariant state available for the given (infinite) energy.
{
Third, algebraic models for experiments set against a background Minkowski space constitute a block-world
formalism}, in which there is essentially no need for evolution of a time-dependent state as a fundamental
part of the formalism (this is not an ontological claim that the world is a 4-dimensional block world,
only a description of the mathematical components of this kind of model).
The Hamiltonian is secondary to the algebraic structure in an algebraic formulation, the commutation
relations of the creation and annihilation operators and the definition of the vacuum vector determine
the results of all observables in the vacuum state and in every state that results from the GNS construction.

The random field is the more natural construction, insofar as it does not require an explicit separation
into positive and negative frequencies.
The random field is also a promising mathematical starting point, insofar as Lie field deformations
that preserve the commutativity of the random field are possible~\cite{MorganJMP}.
The presentation of a Klein-Gordon random field in terms of an algebra of creation and annihilation operators
introduces many of the peculiarities of quantum field theory, but it remains a classical {mathematics of
stochastic signal analysis} at the level of the random field, suggesting possibilities for an interpretation of
quantum field theory, subject to a detailed understanding of the role of fermions and gauge invariance.

\acknowledgments
The author thanks Ken Wharton and a referee for helpful comments.

\end{document}